\documentclass[11pt]{article}

\usepackage[cp1251]{inputenc}

\begin{document}

\begin{center}

{\bf  V.V. Kisel, E.M. Ovsiyuk\footnote{e.ovsiyuk@mail.ru}, V.M. Red'kov\footnote{redkov@dragon.bas-net.by}, N.G. Tokarevskaya \\[3mm]
 EXACT SOLUTIONS FOR A QUANTUM-MECHANICAL PARTICLE  WITH SPIN 1
 AND ADDITIONAL INTRINSIC CHARACTERISTICS  IN A HOMOGENEOUS MAGNETIC
 FIELD}\\[3mm]
 {\em Institute of Physics, National Academy of Sciences of Belarus
 }


\end{center}

\vspace{5mm}

\begin{quotation}

With the use of the general covariant matrix 10-dimensional  Petiau -- Duffin
-- Kemmer formalism in cylindrical coordinates
  exact solutions of the quantum-mechanical
equation for a particle with spin 1 in the presence of an external
homogeneous magnetic field are constructed. Three linearly
independent types of solutions are separated; in each case the formula for
the energy levels has been found.
Within  similar technique  for the
 quantum-mechanical equation
for a particle with spin 1  and additional intrinsic  electromagnetic
characteristics -- polarizability, exact solutions  are found  in the presence of an external
homogeneous magnetic field.

\end{quotation}


\section{Introduction}

The  problem of a quantum-mechanical particle in an  external homogeneous  magnetic field is well-known
in theoretical physics. In fact, only two cases are considered: a scalar (Schr\"{o}dinger's)
 non-relativistic particle with spin 0, and fermions
(non- relativistic Pauli's and relativistic Dirac's) with spin
$1/2$ (the first investigation were \cite{1',2',3',4'}).
In the case of spin 1 particle, the most popular
quantum-mechanical problem is the Coulomb one
\cite{4'}.

In the first part of the paper (Sections {\bf 1 -- 3}),  exact solutions for a ordinary vector particle
will be constructed.
In the second  part (Sections {\bf 4 -- 6}), the exact solutions for a particle with spin 1
and an additional intrinsic electromagnetic parameter (polarizability)
will be constructed explicitly  as well. In principle, these results provide us with  a  possibility for experimental
testing of this characteristics  -- polarizability of the spin 1
particle.

 To treat the problem for an ordinary vector particle
 we take the matrix Petiau -- Duffin --  Kemmer approach
 extended to a general covariant form on the basis of the tetrad formalism (recent
 consideration and   references see e.g., in  \cite{5',6'}).
The main equation in the tetrad form reads \cite{6'}
\begin{eqnarray}
 \left [ \; i\;
\beta^{\alpha}(x) \; ( \partial_{\alpha} + B_{\alpha} - i {e \over
\hbar } A_{\alpha} ) \;- \;
 {Mc\over \hbar}\; \right ] \; \Psi (x) = 0 \; ,
\nonumber
\\
\beta^{\alpha} (x) = \beta^{a} \; e_{(a)}^{\alpha} (x) \; , \qquad
B_{\alpha}(x) = {1 \over 2 } \; J^{ab} \; e_{(a)}^{\beta}
\nabla_{\alpha} e_{(b)\beta} \; ; \label{2.1}
\end{eqnarray}

\noindent $e_{(a)}^{\alpha} (x)$ is a tetrad,  $J^{ab}$ stands for generators for
 10-dimensional representation of the Lorentz group referred to 4-vector and anti-symmetric tensor
  (for brevity  we note
 $Mc / \hbar$ as $M $).
The homogeneous magnetic field  ${\bf B} = (0,0,B)$ corresponds to 4-potential
$
 A^{a} =  (\;  0, {1 \over 2} \; \vec{B} \times \vec{r} \;)$;
 in the cylindric coordinates,  and the last is given by
\begin{eqnarray}
dS^{2} = c^{2} dt^{2} - d r^{2} -
r^{2} \; d\phi^{2} - dz^{2} \;, \qquad   A_{\phi} = - B r^{2} / 2   \; . \label{2.2c}
\end{eqnarray}

Choosing a diagonal cylindric tetrad
\begin{eqnarray}
e_{(0)}^{\alpha} = (1,0,0,0 ) \; , \;\;  e_{(1)}^{\alpha} =
(0,1,0,0 ) \; ,\;
\;\; e_{(2)}^{\alpha} = (0,0, {1 \over r
}, 0 ) \; , \;\; e_{(3)}^{\alpha} = (0,0,0,1 ) \; ,
\nonumber
\label{2.3}
\end{eqnarray}

\noindent after simple calculations,  the main equation (\ref{2.1})  reduces to the form
 \begin{eqnarray}
\left [  i \beta^{0}  \partial_{0}  +  i \beta^{1}
\partial_{r} + i {\beta^{2}  \over  r }
 (  \partial _{\phi} +  {ie B \over 2 \hbar }   r^{2}   + J^{12} ) +
i \beta^{3}  \partial_{z} - M  \right ]  \Psi =
0 \; . \label{2.7'}
\end{eqnarray}

\noindent For brevity we will note $(eB/2  \hbar)$ as $B$. It is better  to choose the matrices  $\beta^{a}$
in the so-called cyclic form, where  the generator  $J^{12}$ has a diagonal structure.
These matrices are  given in \cite{6'}.

\section{Separation of  variables }

With the use of a special substitution (it corresponds to diagonalization of the third projections of momentum $P_{3}$
and angular momentum $J_{3}$ for a particle  with spin 1, specified to the  cylindric tetrad basis)
\begin{eqnarray}
\Psi = e^{-i\epsilon t  }  e^{im\phi}  e^{ikz} \left |
\begin{array}{c}
\Phi_{0} \\
\vec{\Phi} \\
\vec{E} \\
\vec{H}
\end{array} \right |,
\label{3.2}
\end{eqnarray}

\noindent the main equation reads
\begin{eqnarray}
 \left [  \epsilon \beta^{0}   + i\beta^{1}
\partial_{r} - { \beta^{2}  \over  r }
 (   m + Br^{2}  - S_{3} ) - k  \beta^{3}  - M  \right ]  \left | \begin{array}{c}
\Phi_{0} \\
\vec{\Phi} \\
\vec{E} \\
\vec{H}
\end{array} \right |
 = 0  \;;
\nonumber
\end{eqnarray}

\noindent  after calculations  we arrive at the radial system of 10 equations
\begin{eqnarray}
-\hat{b}_{m-1} \;  E_{1} -   \hat{a}_{m+1} \;  E_{3}   -     ik \;  E_{2} = M
\Phi_{0} \; , \nonumber
\\
-i  \hat{b}_{m-1}  \; H_{1}  + i  \hat{a}_{m+1}  \;  H_{3}  +  i \epsilon  \;
E_{2}   = M  \Phi_{2}\;, \nonumber
\\
  i  \hat{a}_{m}   \; H_{2}  +i \epsilon \; E_{1}   -  k\; H_{1} = M  \Phi_{1}\;,
\nonumber
\\
 -i  \hat{b}_{m}  \; H_{2} + i \epsilon  \; E_{3} +  k \; H_{3} = M  \Phi_{3}\;,
\label{3.9a}
\end{eqnarray}
\begin{eqnarray}
    \hat{a}_{m}  \; \Phi_{0} -i  \epsilon  \; \Phi_{1}  = M  E_{1}\;, \qquad
    -i   \hat{a}_{m}   \; \Phi_{2}  + k \;\Phi_{1} = M   H_{1}\;,
\nonumber
\\
   \hat{b}_{m}   \; \Phi_{0}  -i  \epsilon  \;  \Phi_{3}  = M  E_{3}\;, \qquad
i \hat{b}_{m}  \;  \Phi_{2}   -   k \; \Phi_{3}= M  H_{3}\;, \nonumber
\\
-i \epsilon  \;  \Phi_{2}  -  i  k \; \Phi_{0} =M   E_{2}\;,
\qquad
  i   \hat{b}_{m-1}  \; \Phi_{1} -  i  \hat{a}_{m+1}  \;   \Phi_{3}
 = M   H_{2} \; ,
\label{3.9b}
\end{eqnarray}

\noindent where special abbreviations were used for first order differential operators
\begin{eqnarray}
{1 \over \sqrt{2}}  (  {d   \over d r }    +  {m + Br^{2}   \over
r } ) =    \hat{a}_{m}\; ,\qquad  {1 \over \sqrt{2}} (  - {d   \over d r
}    +  {m + Br^{2}   \over r } ) =   \hat{b}_{m} \; . \nonumber
\label{3.8}
\end{eqnarray}

From (\ref{3.9a}) -- (\ref{3.9b}) it follows 4 equations for the components  $\Phi_{a}$
\begin{eqnarray}
(-\hat{b}_{m-1} \; \hat{a}_{m}   -  \hat{a}_{m+1} \hat{b}_{m}   -   k ^{2}  - M^{2} )\;
\Phi_{0} -
  \epsilon  k \;  \Phi_{2}
  \nonumber
  \\
 +   i  \epsilon  \; ( \;  \hat{b}_{m-1}  \Phi_{1}    +  \hat{a}_{m+1}  \Phi_{3} \;  ) = 0  \; ,
\nonumber
\\
( \; -    \hat{b}_{m-1}  \hat{a}_{m}   -   \hat{a}_{m+1}  \hat{b}_{m}  +   \epsilon ^{2}
-M^{2} \; )\;  \Phi_{2} +  \epsilon k \; \Phi_{0}
\nonumber
\\
-  ik \;( \;
\hat{b}_{m-1} \Phi_{1}      +  \hat{a}_{m+1}   \Phi_{3} \; )         = 0 \;,
\nonumber
\\
( \; -   \hat{a}_{m}  \hat{b}_{m-1}    +  \epsilon^{2}   - k^{2}      -   M
^{2} \;  ) \; \Phi_{1}   + \hat{a}_{m}  \hat{a}_{m+1}  \;   \Phi_{3}
\nonumber
\\
+  i
\epsilon \;  \hat{a}_{m}  \; \Phi_{0} + i k  \; \hat{a}_{m}   \Phi_{2}     = 0
\; , \nonumber
\\
(\; -  \hat{b}_{m} \hat{a}_{m+1}  + \epsilon ^{2}    - M^{2} - k^{2} \;  )\;
\Phi_{3} +  \hat{b}_{m} \hat{b}_{m-1}   \Phi_{1}   +
\nonumber
\\
+  i \epsilon \; \hat{b}_{m}
\Phi_{0}      +     ik\;  \hat{b}_{m}    \Phi_{2}      = 0 \; ;
 \label{3.10}
\end{eqnarray}

\section{General analysis of the radial  equations}

Eqs.  (\ref{3.10}) can be transformed to the form
\begin{eqnarray}
[  -\hat{b}_{m-1}   \hat{a}_{m}   -  \hat{a}_{m+1} \hat{b}_{m}   + \epsilon^{2}    -
M^{2} -   k ^{2}    ] \;
  (  k  \Phi_{0}  + \epsilon   \Phi_{2}  )
 = 0 \; ,
\nonumber
\\
\; [   -\hat{b}_{m-1}  \hat{a}_{m}   -  \hat{a}_{m+1} \hat{b}_{m} + \epsilon^{2} -
k^{2} - M^{2}   ]  ( \epsilon   \Phi_{0} +  k  \Phi_{2} )
\nonumber
\\
= (\epsilon^{2} - k^{2})  [ ( \epsilon   \Phi_{0} +  k
\Phi_{2} )
  -   (  i \hat{b}_{m-1}  \Phi_{1}    +  i \hat{a}_{m+1}  \Phi_{3}   ) ]  \;;
\label{5.2a}
\\[2mm]
(  -   \hat{a}_{m}  \hat{b}_{m-1}    +  \epsilon^{2}   - k^{2}      -   M
^{2}   )  \Phi_{1}   + \hat{a}_{m}  \hat{a}_{m+1}  \;   \Phi_{3}    +    i
\epsilon   \hat{a}_{m}   \Phi_{0} + i k   \hat{a}_{m}   \Phi_{2}     = 0
\; , \nonumber
\\
( -  \hat{b}_{m} \hat{a}_{m+1}  + \epsilon ^{2}    - M^{2} - k^{2}   )
\Phi_{3} +  \hat{b}_{m} \hat{b}_{m-1}   \Phi_{1}    +     i \epsilon  \hat{b}_{m}
\Phi_{0}      +     ik   \hat{b}_{m}    \Phi_{2}      = 0 \; .
\nonumber
\\
\label{5.2b}
\end{eqnarray}

\noindent
Let us introduce new variables
\begin{eqnarray}
F(r) =   k \; \Phi_{0}(r)  + \epsilon \; \Phi_{2}(r)\;  , \qquad
 G (r) =  \epsilon \;  \Phi_{0}(r)  +  k \; \Phi_{2} (r) \; ,
 \label{D}
 \end{eqnarray}

\noindent then Eqs.  (\ref{5.2a}) -- (\ref{5.2b})  read
\begin{eqnarray}
[\; -\hat{b}_{m-1} \; \hat{a}_{m}   -  \hat{a}_{m+1} \hat{b}_{m}   + \epsilon^{2}    -
M^{2} -   k ^{2}  \; ] \;F
   = 0 \; ,
\nonumber
\\
\; [ \; -\hat{b}_{m-1}  \hat{a}_{m}   -  \hat{a}_{m+1} \hat{b}_{m}    - M^{2}\;   ]\; G
\nonumber
\\
=
  - (\epsilon^{2} - k^{2} ) \;  (  i \hat{b}_{m-1}  \Phi_{1}    +  i \hat{a}_{m+1}  \Phi_{3}   )\; ]  \;,
\label{5.4a}
\end{eqnarray}
\begin{eqnarray}
( \; -   \hat{a}_{m}  \hat{b}_{m-1}    +  \epsilon^{2}   - k^{2}      -   M
^{2} \;  ) \; \Phi_{1}   + \hat{a}_{m}  \hat{a}_{m+1}  \;   \Phi_{3}    +    i
\hat{a}_{m} \; G    = 0 \; , \nonumber
\\
(\; -  \hat{b}_{m} \hat{a}_{m+1}  + \epsilon ^{2}    - M^{2} - k^{2} \;  )\;
\Phi_{3} +  \hat{b}_{m} \hat{b}_{m-1}   \Phi_{1}    +     i  \; \hat{b}_{m} \;  G =
0 \; . \label{5.4b}
\end{eqnarray}

For  equations (\ref{5.4b}), let us multiply the first one (from the left)  by  $\hat{b}_{m-1}$ and
the second one by the  $\hat{a}_{m+1}$ , that results in
\begin{eqnarray}
- \hat{b}_{m-1}  \hat{a}_{m}  ( \hat{b}_{m-1}  \Phi_{1} )     + ( \epsilon^{2}   -
k^{2}      -   M ^{2} )   ) (\hat{b}_{m-1} \Phi_{1} )
\nonumber
\\
+ \hat{b}_{m-1} \hat{a}_{m}
(\hat{a}_{m+1}    \Phi_{3} )   +    i  \hat{b}_{m-1} \hat{a}_{m}  G    = 0 \; ,
\nonumber
\\
- \hat{a}_{m+1}  \hat{b}_{m} (\hat{a}_{m+1} \Phi_{3})  + (\epsilon ^{2}    - M^{2} -
k^{2} ) (\hat{a}_{m+1} \Phi_{3})
\nonumber
\\
+ \hat{a}_{m+1}  \hat{b}_{m} (\hat{b}_{m-1}   \Phi_{1}) +
i  \; \hat{a}_{m+1} \hat{b}_{m} \;  G     = 0 \; .
\label{5.5b}
\end{eqnarray}

\noindent Again, let us introduce  two new  field variables
\begin{eqnarray}
 \hat{b}_{m-1}  \Phi_{1}  = Z_{1}\;  , \qquad \hat{a}_{m+1}    \Phi_{3}  = Z_{3} \; ;
\label{5.6}
\end{eqnarray}

\noindent Eqs.  (\ref{5.5b}) read as follows
\begin{eqnarray}
- \hat{b}_{m-1}  \hat{a}_{m}  Z_{1}     + ( \epsilon^{2}   - k^{2}      - M
^{2}    ) Z_{1}  + \hat{b}_{m-1} \hat{a}_{m}  Z_{3}    +    i  \hat{b}_{m-1} \hat{a}_{m}
G    = 0 \; , \nonumber
\\
- \hat{a}_{m+1}  \hat{b}_{m}  Z_{3}   + (\epsilon ^{2}    - M^{2} - k^{2} )
Z_{3} + \hat{a}_{m+1}  \hat{b}_{m}  Z_{1}     +     i  \; \hat{a}_{m+1} \hat{b}_{m} \;  G
= 0 \; . \label{5.7}
\end{eqnarray}

\noindent With the help of new  functions  $f(r), g(r)$
\begin{eqnarray}
Z_{1} = {f+g\over 2}\; , \;\; Z_{3}= {f-g \over 2} \; , \qquad
Z_{1}+ Z_{3} = f\; ,  \;\;  Z_{1}- Z_{3} = g \; ; \label{5.8}
\end{eqnarray}

\noindent the system (\ref{5.7})  is transformed to the following form
\begin{eqnarray}
- \hat{b}_{m-1}  \hat{a}_{m} \;  g     +
 ( \epsilon^{2}   - k^{2}      -   M ^{2}    ) {f+g\over 2}
   +    i  \hat{b}_{m-1} \hat{a}_{m}  G    = 0 \; ,
\nonumber
\\
 \hat{a}_{m+1}  \hat{b}_{m} \;  g   + (\epsilon ^{2}    - M^{2} - k^{2} )  {f-g \over 2}
   +     i  \; \hat{a}_{m+1} \hat{b}_{m} \;  G     = 0 \; .
\label{5.9}
\end{eqnarray}

\noindent Combining these equations we get
\begin{eqnarray}
[ - \hat{b}_{m-1}  \hat{a}_{m}    -  \hat{a}_{m+1}  \hat{b}_{m}  +  \epsilon^{2}   -
k^{2}      -   M ^{2}     ]  g
      +    i (  \hat{b}_{m-1} \hat{a}_{m}  -   \; \hat{a}_{m+1} \hat{b}_{m}  )  G    = 0 \; ,
\nonumber
\\
( - \hat{b}_{m-1}  \hat{a}_{m}       + \hat{a}_{m+1}  \hat{b}_{m}  )   g +
 ( \epsilon^{2}   - k^{2}      -   M ^{2}    ) f
   +    i  ( \hat{b}_{m-1} \hat{a}_{m}  +  \hat{a}_{m+1} \hat{b}_{m} )  G    = 0 \; .
\nonumber
\\
\label{5.10}
\end{eqnarray}

In turn,   Eqs.  (\ref{5.4a})  can be  presented  as
\begin{eqnarray}
(\; -\hat{b}_{m-1} \; \hat{a}_{m}   -  \hat{a}_{m+1} \hat{b}_{m}   + \epsilon^{2}    -
M^{2} -   k ^{2}  \; ) \;F
   = 0 \; ,
\nonumber
\\
( \; -\hat{b}_{m-1}  \hat{a}_{m}   -  \hat{a}_{m+1} \hat{b}_{m}    - M^{2}\;   )\; G =
  - i (\epsilon^{2} - k^{2} ) \;  f   \; .
\label{5.11}
\end{eqnarray}

\noindent
Further, with  the use of identities
\begin{eqnarray}
-\hat{b}_{m-1} \; \hat{a}_{m}   -  \hat{a}_{m+1} \hat{b}_{m} =  \Delta  \; , \qquad -
\hat{b}_{m-1} \; \hat{a}_{m}   +  \hat{a}_{m+1} \hat{b}_{m} = 2 B  \;  \label{5.12}
\end{eqnarray}

\noindent Eqs.  (\ref{5.11})  and  (\ref{5.10})   can be written  as follows
\begin{eqnarray}
( \Delta    + \epsilon^{2}    - M^{2} -   k ^{2}   ) \;F
   = 0 \; ,
\nonumber
\\
  \Delta \;  G  =  M^{2}     G      - i  (\epsilon^{2} - k^{2} ) \; f     \; ,
\nonumber
\\
(  \Delta   +  \epsilon^{2}   - k^{2}      -   M ^{2}     )\;  g =
           2i B  \; G     \; ,
\nonumber
\\
 ( \epsilon^{2}   - k^{2}      -   M ^{2}    ) \; f
   -     i \Delta \; G   +   2B \;     g   = 0 \; .
\label{5.13}
\end{eqnarray}

\noindent
With the help of the second equation, from the forth one it follows
\begin{eqnarray}
              \; f =
   -     i    \;    G        +  { 2B  \over  M ^{2} } \;     g    \; .
\label{5.14}
\end{eqnarray}

\noindent Now, one excludes the function  $f$ in the second equation in
(\ref{5.13}) and gets
\begin{eqnarray}
 ( \Delta + \epsilon^{2} - k^{2}  -  M^{2}  ) \;    G  =     - i  (\epsilon^{2} - k^{2} )
   { 2B  \over  M ^{2} } \;     g       \;.
\label{5.15}
\end{eqnarray}

Thus, the general problem is reduced  to the system of four equations
\begin{eqnarray}
( \Delta    + \epsilon^{2}    - M^{2} -   k ^{2}   ) \;F
   = 0 \; ,
   \nonumber
   \\
              \; f =
   -     i    \;    G        +  { 2B  \over  M ^{2} } \;     g    \; ,
\nonumber
\\
( \;  \Delta   +  \epsilon^{2}   - k^{2}      -   M ^{2}  \; )\;
g =
           2i B  \; G     \; ,
           \nonumber
           \\
 (\;  \Delta + \epsilon^{2} - k^{2}  -  M^{2} \;  ) \;    G  =     - 2i B \; {\epsilon^{2} - k^{2} \over  M ^{2} }
      \;     g       \; .
\label{5.16}
\end{eqnarray}

The structure of this system allows to separate an evident linearly independent solution
as follows
\begin{eqnarray}
f(r)=0 \; ,\qquad g(r)=0\;, \qquad H(r)= 0 \; ,
\nonumber\\
F(r) \neq 0 \; , \qquad  (  \Delta  -k^{2} -M^{2}+  \epsilon ^{2}
) \; F  = 0 \; .\label{3.32}
\end{eqnarray}

\noindent Corresponding functions and energy spectrum are known.
We are to solve the system of two last equations in (\ref{5.16}); in matrix form it reads
(let $\gamma = (\epsilon^{2} - k^{2}) / M ^{2} $)
\begin{eqnarray}
( \Delta    + \epsilon^{2}    - M^{2} -   k ^{2}   ) \left |
\begin{array}{c}
g \\
G
\end{array} \right | =
\left | \begin{array}{cc}
0  & 2iB \\
-2iB \gamma & 0
\end{array} \right |
  \left | \begin{array}{c}
g  \\
G
\end{array} \right | \; .
\label{5.17}
\end{eqnarray}

\noindent Let us  construct the transformation changing the matrix on the right to a diagonal form
\begin{eqnarray}
( \Delta    + \epsilon^{2}    - M^{2} -   k ^{2}   )  \left |
\begin{array}{c}
g '\\
G '
\end{array} \right | =
\left | \begin{array}{cc}
\lambda_{1}   & 0 \\
0   & \lambda_{2}
\end{array} \right |
  \left | \begin{array}{c}
g '\\
G'
\end{array} \right |,
\nonumber
\\
\left | \begin{array}{c}
g '\\
G'
\end{array} \right | = S \; \left | \begin{array}{c}
g \\
G
\end{array} \right | \; , \qquad
S = \left | \begin{array}{cc}
s_{11} & s_{12} \\
s_{21} & s_{22}
\end{array} \right | .
\label{5.18}
\end{eqnarray}

\noindent The problem reduces to
  linear systems
\begin{eqnarray}
\left \{ \begin{array}{l}
- \lambda_{1} \; s_{11} - 2iB\gamma \; s_{12} = 0 \; ,\\
2iB \; s_{11} - \lambda_{1}\; s_{12} = 0 \; ,
\end{array} \right. \qquad
\left \{ \begin{array}{l}
- \lambda_{2} \; s_{21} - 2iB\gamma \; s_{22} = 0 \; ,\\
2iB \; s_{21} - \lambda_{2}\; s_{22} = 0 \; .
\end{array} \right.
\nonumber
\end{eqnarray}

\noindent The values  of $\lambda_{1}$ and $ \lambda_{2}$
are given by
\begin{eqnarray}
\lambda_{1} = + 2B \sqrt{\gamma} \; , \qquad \lambda_{2} = -
2B \sqrt{\gamma} \; ,
\nonumber
\\
i \; s_{11} -   \sqrt{\gamma} \; s_{12} = 0 \; , \qquad i \;
s_{21} +  \sqrt{\gamma}\; s_{22} = 0 \; ,
\nonumber
\\
s_{12}=1,  \; s_{22}=1 \;, \qquad
  S = \left | \begin{array}{rr}
-i \; \sqrt{\gamma}   & 1  \\
+i \; \sqrt{\gamma}   & 1
\end{array} \right |.
\label{5.20b}
\end{eqnarray}

In the new (primed) basis, Eqs.  (\ref{5.17})  take the form of two separated differential equations
\begin{eqnarray}
 \left (  \; \Delta   +
  \epsilon^{2}   - k^{2}      -   M ^{2}  -  2B \; \sqrt{\gamma}   \;  \right  )  \; g' =0 \; ,
\nonumber
\\
   \left ( \;  \Delta   +  \epsilon^{2}   - k^{2}      -   M ^{2}  + 2B \; \sqrt{\gamma}  \;  \right  )  \; G' =0 \; ;
\label{5.22a}
\end{eqnarray}

Recalling the meaning of $\Delta$, let us specify the second order differential equation
\begin{eqnarray}
   \left (
   {d^{2} \over dr^{2}}  + {1 \over r}{d \over d r} -
{(m+Br^{2})^{2} \over r^{2}}     + \lambda^{2}  \right  )
\varphi  (r) =0 \; , \nonumber
\\
\lambda^{2} =  \epsilon^{2}   - k^{2}      -   M ^{2}  \pm
 2B \; \sqrt{\gamma}    , \qquad \sqrt{\gamma} =  {\sqrt{\epsilon^{2} - k^{2} }\over  M  } \; .
\label{dif}
\end{eqnarray}

It is convenient to introduce a new variable $ x = B r^{2}$, then Eq. (\ref{dif})
 reads\footnote{For definiteness let us consider $B$ to be positive, which does not affect the generality of the analysis.
So, to infinite values of  $r$ correspond infinite and positive values of  $x$.}
\begin{eqnarray}
x{d^{2}\varphi\over dx^{2}}+ {d\varphi\over dx}-\left(
{m^{2}\over4 x}+{x\over 4}+{m\over 2}-{\lambda^{2}\over
4B}\right)\varphi=0\,. \label{5.23}
\end{eqnarray}

\noindent With the substitution $ \varphi (x) = x^{A} e^{-Cx} f (x)
\; , $  for $f(x)$ we get
\begin{eqnarray}
x{d^{2}f\over dx^{2}}\,+  \left(2A+1-2Cx\right){df\over dx}
\nonumber
\\
+
 \left[{A^{2}-m^{2}/4\over x}+ (C^{2}-{1\over 4} )x-2AC-C-{m\over 2}+{\lambda^{2}\over 4B}\right]f=0\,.
\nonumber \label{5.24}
\end{eqnarray}

\noindent When  $A, C$ are taken as
$ A= +  \mid m \mid /2 \; , \;  C= + 1 /2 $
the previous equation becomes simpler
\begin{eqnarray}
x{d^{2}R\over dx^{2}}\,+\left(2A+1-x\right){dR\over dx}-
\left(A+{1\over 2}+{m\over 2}-{\lambda^{2}\over 4B}\right)R=0\, ,
\nonumber
\end{eqnarray}

\noindent which is of confluent hypergeometric type
\begin{eqnarray}
x \;  Y '' +( \gamma -x) Y'  - \alpha Y =0 \; , \nonumber \\
\alpha= {\mid m \mid \over 2} +{1\over 2}+{m\over
2}-{\lambda^{2}\over 4B}\,, \qquad \gamma= \mid m \mid +1\,.
\nonumber
\end{eqnarray}

\noindent To obtain polynomials we must impose  an additional condition
$ \alpha = - n\;; $ which  provides us with the following quantization rule
for
$\lambda^{2}$
\begin{eqnarray}
\lambda^{2} = 4B \; ( n + {1 \over 2} + { \mid m \mid + m \over 2}
)\; . \label{5.25}
\end{eqnarray}

Thus, we have arrived at two formulas for the energy
\begin{eqnarray}
 \sqrt{\epsilon^{2} - k^{2}}  =  { +B  +   \sqrt{B^{2} +M^{2} (M^{2} + \lambda^{2})} \over M}  \; ,
 \nonumber
 \\
\sqrt{\epsilon^{2} - k^{2}}  =  { -B  +   \sqrt{B^{2} +M^{2}
(M^{2} + \lambda^{2})} \over M} \;  .
\label{5.28}
\end{eqnarray}

\noindent In turn, the energy spectrum for the case (\ref{3.32}) is given by
\begin{eqnarray}
\epsilon^{2} = M^{2} + k^{2} + \lambda ^{2} \; \label{5.29}
\end{eqnarray}

Thus,  on the base of the use of general covariant formalism in the Petiau -- Duffin -- Kemmer theory
 of  the vector particle,  the exact solutions for such a particle  are constructed in the
 presence of an external homogeneous magnetic
 field.
 There are separated three types of linearly independent solutions, and the corresponding energy spectra are found.

\section{ On a spin 1 particle with intrinsic  structure  -- polarizability
}

\hspace{5mm} In  \cite{1}--\cite{7}, it was described  a generalized
equation for spin 1 particle possessing in addition to electric
charge the special electromagnetic characteristics named
polarizability. In the frame of the first order relativistic wave
equations, such a  particle  requires a 15-dimensional wave
function, consisting of a 4-vector $\Phi_{a}(x)$,
 4-tensor $\Phi_{ab}(x)$, and subsidiary  scalar and 4-vector fields,  $C(x)$ and  $C_{a}(x)$.


 To treat the problem we take the matrix  approach in the theory of the generalized $S=1$ particle
 extended to a general covariant form on the base of  tetrad formalism (recent
 consideration, notation  and list of references see in \cite{13, 14}). The use of cylindric tetrad  permits
 to  take account of the cylindric  symmetry of  the problem.
The main equation in tetrad form is \cite{6}
\begin{eqnarray}
\left [  \Gamma^{0}  \partial_{0}  + \Gamma^{1}
\partial_{r} + {1 \over  r } \; \Gamma^{2}  (  \partial
_{\phi}  + \; {ie B \over 2 \hbar } \;  r^{2}  \; + J^{12} ) + \Gamma^{3}   \partial_{z} - M  \right ]
 \Psi  = 0 \; . \label{2.7}
\end{eqnarray}

It is better  to choose the matrices  $\beta^{a}$ in the so-called
cyclic form, where  the generator  $J^{12}$ has a diagonal
structure. These matrices  $\Gamma^{a}$ are given in \cite{6'}.

\section{Separation of variables  }

With the use of special substitution
\begin{eqnarray}
\Psi = \{ \; C,\;  C_{0}, \; \vec{C}, \; \Phi_{0}, \vec{\Phi},
\; \vec{E}, \; \vec{H} \; \} \; , \qquad C (x) = e^{-i \epsilon
t/\hbar   }  e^{ikz}  e^{im \phi } \; C(r) \; ,
\nonumber\\
C_{0}  = e^{-i \epsilon t/ \hbar   } e^{ikz}  e^{im \phi }
\; C_{0}(r) \; , \qquad  \vec{C}  = e^{-i \epsilon t/ \hbar   }
 e^{ikz}  e^{im \phi }  \left | \begin{array}{c} C_{1} (r)
\\ C_{2} (r) \\ C_{3} (r)
\end{array} \right | ,
\nonumber\\
\Phi_{0}  = e^{-i \epsilon t/ \hbar   }  e^{ikz}  e^{im \phi
}  \Phi_{0}(r) \; , \qquad \vec{\Phi}  = e^{-i \epsilon t/
\hbar   }  e^{ikz}  e^{im \phi }  \left | \begin{array}{c}
\Phi_{1} (r) \\ \Phi_{2} (r) \\  \Phi_{3} (r)
\end{array} \right | ,
\nonumber\\
\vec{E}  = e^{-i \epsilon t/ \hbar   }  e^{ikz}  e^{im \phi
}  \left | \begin{array}{c} E_{1} (r) \\ E_{2} (r) \\  E_{3} (r)
\end{array} \right | , \;\;\;
\vec{H}  = e^{-i \epsilon t/ \hbar   }  e^{ikz}  e^{im \phi
}  \left | \begin{array}{c} H_{1} (r) \\ H_{2} (r) \\  H_{3} (r)
\end{array} \right | ,
\label{3.4}
\end{eqnarray}

\noindent after calculations  we arrive at the radial system of 15
equations
\begin{eqnarray}
 - i \epsilon  \;C_{0}
-\hat{b}_{m-1} \; C_1- \hat{a}_{m+1} \; C_{3}- ik\; C_{2}= M \; C
\; , \label{3.7a}
\\
-\hat{b}_{m-1} \; E_{1}-\hat{a}_{m+1} \; E_{3}-ik \; E_{2}= M
\;C_{0}\; ,
\nonumber\\
i\epsilon \;E_{1} + i\hat{a}_{m} \; H_{2}- ik \; H_{1} = M \;C_{1}
\; ,
\nonumber\\
i\epsilon  \;E_{2}- i\hat{b}_{m-1} \; H_{1} + i\hat{a}_{m+1} \;
H_{3} = M \;C_{2}\; ,
\nonumber\\
 i\epsilon  \;E_{3}-i\hat{b}_{m} \; H_{2}+ k \; H_{3}= M \;C_{3},
\label{3.7b}
\\
 - i\epsilon \;\sigma C - \hat{b}_{m-1}\;
 E_{1}- \hat{a}_{m+1} \; E_{3}- ik \; E_{2}= M \;\Phi_{0} \; ,
\nonumber\\
i\epsilon \;E_{1}- \sigma\; \hat{a}_{m}\; C + i\; \hat{a}_{m}\;
H_{2}-k \; H_{1} = M \;\Phi_{1} \; ,
\nonumber\\
i\epsilon \; E_{2}- i\hat{b}_{m-1} \; H_{1} + i \hat{a}_{m+1} \;
H_{3}+ i\;k\sigma \; C = M \;\Phi_{2}\; ,
\nonumber\\
i\epsilon \;E_{3} - \sigma \;\hat{b}_{m} \; C- i\hat{b}_{m} \;
H_{2} +k \; H_{3}= M \;\Phi_{3}\; , \label{3.7c}
\\
 - i\epsilon \;\Phi_{1} + \hat{a}_{m}\; \Phi_{0} = M \;E_{1} \; ,
\qquad -i\epsilon \;\Phi_{2}- ik \;\Phi_{0}= M \;E_{2}\; ,
\nonumber\\
-i\epsilon \;\Phi_{3}+ \hat{b}_{m}\;\Phi_{0} =M \;E_{3}\; , \qquad
-i\hat{a}_{m} \; \Phi_{2} + k \; \Phi_{1}=M \;H_{1}\; ,
\nonumber\\
i\hat{b}_{m-1} \; \Phi_{1} - i\hat{a}_{m+1} \; \Phi_{3} = M
\;H_{2} \; ,\qquad i\hat{b}_{m} \; \Phi_{2} - k \; \Phi_{3}=M
\;H_{3} \;. \label{3.7d}
\end{eqnarray}

\section{Solution  of the radial system
}

With the use of  (\ref{3.7b}), Eqs.   (\ref{3.7c}) give
\begin{eqnarray}
C_{0}= \Phi_{0}+ \;i\; {\epsilon \;\sigma \over  M}\; C \; ,
\qquad C_{1}= \Phi_{1} +\; {\sigma \over  M}\; \hat{a}_{m} C \; ,
\nonumber\\
C_{2}=\Phi_{2}-\;i\;{k\;\sigma \over  M}\;  C \; , \qquad C_{3}=
\Phi_{3}+\;{\sigma \over  M}\; \hat{b}_{m} C \; . \label{3.8}
\end{eqnarray}

\noindent Substituting these  formulas for  $C_{a}$ into
(\ref{3.7a})
\begin{eqnarray}
-\;i\; \epsilon   (\Phi_{0}+\;i\; {\epsilon \;\sigma \over M}\; C
) \; -\; \hat{b}_{m-1}\;  (\Phi_{1}+\;{\sigma \over  M}\;
\hat{a}_{m} \; C \;  )
\nonumber\\
-\;\hat{a}_{m+1} (\Phi_{3} \; +\;{\sigma \over  M}\; \hat{b}_{m}\;
C  )- \;ik \;  ( \Phi_{2} \; - \;i\;{k\;\sigma \over  M}\; C  )= M
\; C \; , \nonumber
\end{eqnarray}

\noindent   we further get
\begin{eqnarray}
M(\hat{b}_{m-1} \; \Phi_{1} + \hat{a}_{m+1} \; \Phi_{3}) =
  - iM \;( \epsilon  \Phi_{0} +
  k  \Phi_{2})
\nonumber\\
+\sigma  (  - \hat{b}_{m-1}  \hat{a}_{m} -\hat{a}_{m+1}
\hat{b}_{m} +\epsilon ^{2}-k^{2}  )C  -  M ^{2}  C \; .
\label{3.9}
\end{eqnarray}

\noindent This equation will be required below.

Note that  Eqs. (\ref{3.7c}) and (\ref{3.7d})  include  the main
field variables, 4-vector and 4-tensor, and also the  scalar  $C$
obeying Eq. (\ref{3.9})
\begin{eqnarray}
 - i\epsilon \;\sigma\; C -\hat{b}_{m-1}
 \; E_{1}-\hat{a}_{m+1} \; E_{3}-ik \; E_{2}=M \;\Phi_{0} \; ,
\nonumber\\
i\epsilon  \;E_{1}- \sigma\; \hat{a}_{m}\; C + i\; \hat{a}_{m} \;
H_{2} -k \; H_{1} = M \;\Phi_{1} \; ,
\nonumber\\
i\epsilon  \;E_{2}- i\hat{b}_{m-1} \; H_{1}+\hat{a}_{m+1} \;
H_{3}+ i\;k\sigma \;C = M \;\Phi_{2}\; ,
\nonumber\\
i\epsilon  \;E_{3} - \sigma \;\hat{b}_{m} \; C-i\hat{b}_{m} \;
H_{2}+ k \; H_{3} =M \;\Phi_{3} \; , \label{3.10a}
\\
 -i\; \epsilon  \;\Phi_{1}+ \hat{a}_{m} \;\Phi_{0}= M \;E_{1}\; ,
\qquad -i\; \epsilon  \;\Phi_{2}- ik \;\Phi_{0}= M \;E_{2} \; ,
\nonumber\\
-i\; \epsilon \;\Phi_{3} + \hat{b}_{m} \;\Phi_{0}= M \;E_{3} \; ,
\qquad -i\hat{a}_{m} \; \Phi_{2} + k \; \Phi_{1}=M \;H_{1} \; ,
\nonumber\\
i\hat{b}_{m-1} \; \Phi_{1}- i\hat{a}_{m+1} \; \Phi_{3}=M \;H_{2}
\; , \qquad i\hat{b}_{m} \; \Phi_{2}-k \; \Phi_{3}=M \;H_{3}\; .
\label{3.11}
\end{eqnarray}

By means of  (\ref{3.11}), we are to eliminate tensor  components
in  (\ref{3.10a}). Then we obtain two equations
\begin{eqnarray}
- \;i\;  \epsilon\; \sigma M  \; C +   (- \hat{b}_{m-1} \;
\hat{a}_{m} - \hat{a}_{m+1} \; \hat{b}_{m}- k^{2}- M^{2} )\Phi_{0}
\nonumber\\
+i\; \epsilon   (\hat{b}_{m-1} \; \Phi_{1}+\hat{a}_{m+1} \;
\Phi_{3}+i k  \; \Phi_{2}  )=0 \; , \label{3.12a}
\end{eqnarray}
\begin{eqnarray}
ik \sigma M \; C +   (  \epsilon^{2} - \hat{b}_{m-1} \;
\hat{a}_{m} - \hat{a}_{m+1} \; \hat{b}_{m} - M^{2}  ) \; \Phi_{2}
\nonumber\\
-i k  ( i  \epsilon \Phi_{0} + \hat{b}_{m-1} \; \Phi_{1} +
\hat{a}_{m+1} \;  \Phi_{3}  ) =0 \; . \label{3.12b}
\end{eqnarray}

\noindent Multiplying the first one  (\ref{3.12a})  by  $+ik$, and
the second  one (\ref{3.12b})  by  $i\epsilon$,  and summing  the
results, we get
\begin{eqnarray}
 ( -\hat{b}_{m-1}\hat{a}_{m}-\hat{a}_{m+1}\hat{b}_{m}-k^{2}
-M^{2}+  \epsilon ^{2}  ) \;
 (k\; \Phi_{0}  + \epsilon\; \Phi_{2} ) = 0 \; .
\label{3.13}
\end{eqnarray}

In the same manner, combining Eqs.  (\ref{3.12a})--(\ref{3.12b})
with  other coefficients, we arrive at
\begin{eqnarray}
( - \hat{b}_{m-1}  \hat{a}_{m} \;  -  \hat{a}_{m+1}   \hat{b}_{m}
- M^{2} )\;
 (  \epsilon \; \Phi_{0}+ k \Phi_{2}  )
\nonumber \\
 =  - i ( \epsilon^{2}-k^{2} ) \; ( \hat{b}_{m-1}
\Phi_{1}  + \hat{a}_{m+1}   \Phi_{3}) + i \sigma  M \; ( \epsilon
^{2}-k^{2} )\; C\; . \label{3.16}
\end{eqnarray}

Thus, two second order equations have been found
\begin{eqnarray}
 ( -\hat{b}_{m-1}\hat{a}_{m}-\hat{a}_{m+1}\hat{b}_{m}-k^{2}
-M^{2}+  \epsilon ^{2}   ) \;
 (k\; \Phi_{0}  + \epsilon\; \Phi_{2} ) = 0 \; ,
\label{3.17a}
\\
( - \hat{b}_{m-1}  \hat{a}_{m} \;  -  \hat{a}_{m+1}   \hat{b}_{m}
- M^{2} )\;
 (  \epsilon \; \Phi_{0}+ k \Phi_{2}  )
\nonumber\\
=  - i ( \epsilon^{2}-k^{2} ) \; ( \hat{b}_{m-1}  \Phi_{1}  +
\hat{a}_{m+1}   \Phi_{3}) + i \sigma  M \; ( \epsilon  ^{2}-k^{2}
)\; C\; .\label{3.17b}
\end{eqnarray}

Now, let us turn to equations in  (\ref{3.10a}),  containing
functions  $m\Phi_{1}$  and  $m\Phi_{3}$:
\begin{eqnarray}
(- \hat{a}_{m} \hat{b}_{m-1}   + \epsilon ^{2} -k^{2}    - M^{2} )
\;\Phi_{1}
        \nonumber\\
    + \hat{a}_{m}\hat{a}_{m+1} \; \Phi_{3}
    +i  \hat{a}_{m} \; (   \epsilon  \;\Phi_{0}  + k \Phi_{2}  )  - M \sigma\; \hat{a}_{m}\; C
  =  0  \; ;
\label{3.18a}
\end{eqnarray}

\noindent and
\begin{eqnarray}
( - \hat{b}_{m} \hat{a}_{m+1}
 + \epsilon^{2} -k^{2}   - M^{2}  ) \;\Phi_{3}
 \nonumber\\
 +   \hat{b}_{m} \hat{b}_{m-1} \; \Phi_{1} +    i \hat{b}_{m} \;  (\epsilon  \Phi_{0} +  k  \Phi_{2} )
   - M \sigma  \;\hat{b}_{m} C
  = 0  \; ,
\label{3.18b}
\end{eqnarray}

\noindent In two last equations, (\ref{3.18a} and  \ref{3.18b}),
multiplying the the first one by   $\hat{b}_{m-1}$  (from the left)
and the second one  by $\hat{a}_{m+1}$  (from the left), we produce
\begin{eqnarray}
(- \hat{b}_{m-1} \hat{a}_{m}    + \epsilon ^{2} -k^{2}    - M^{2}
) \;\hat{b}_{m-1} \Phi_{1}
       \nonumber\\
    + \hat{b}_{m-1} \hat{a}_{m}\hat{a}_{m+1} \; \Phi_{3}
    +i  \hat{b}_{m-1}\hat{a}_{m} \; (   \epsilon  \;\Phi_{0}  + k \Phi_{2}  )  - M \sigma\; \hat{b}_{m-1}\hat{a}_{m}\; C
  =  0  \; ,
\label{3.19}
\\
( - \hat{a}_{m+1} \hat{b}_{m}
 + \epsilon^{2} -k^{2}   - M^{2}  ) \; \hat{a}_{m+1} \Phi_{3}
 \nonumber\\
 +   \hat{a}_{m+1} \hat{b}_{m} \hat{b}_{m-1} \; \Phi_{1} +    i \hat{a}_{m+1} \hat{b}_{m} \;
  (\epsilon  \Phi_{0} +  k  \Phi_{2} )
   - M \sigma  \; \hat{a}_{m+1} \hat{b}_{m} C
  = 0  \; .
\label{3.19b}
\end{eqnarray}

\noindent It is better to introduce new field variables
\begin{eqnarray}
F(r) =   k \; \Phi_{0}   + \epsilon \; \Phi_{2} \;  , \qquad
 G (r) =  \epsilon \;  \Phi_{0}  +  k \; \Phi_{2}  \; ,
\nonumber\\
 \hat{b}_{m-1}  \Phi_{1}  = Z_{1}\;  , \qquad \hat{a}_{m+1}    \Phi_{3}  = Z_{3} \; ;
\label{3.20}
\end{eqnarray}

\noindent then the system  (\ref{3.19})--(\ref{3.19b}) reads
\begin{eqnarray}
(- \hat{b}_{m-1} \hat{a}_{m}    + \epsilon ^{2} -k^{2}    - M^{2}
) \; Z_{1}
        \nonumber\\
    + \hat{b}_{m-1} \hat{a}_{m} Z_{3}
    +i  \hat{b}_{m-1}\hat{a}_{m} \; G  - M \sigma\; \hat{b}_{m-1}\hat{a}_{m}\; C
  =  0  \; ,
\label{3.21a}
\\
( - \hat{a}_{m+1}\hat{b}_{m}
 + \epsilon^{2} -k^{2}   - M^{2}  ) \; Z_{3}
 \nonumber\\
 +   \hat{a}_{m+1} \hat{b}_{m} Z_{1} +    i \hat{a}_{m+1} \hat{b}_{m} \;
 G
   - M \sigma  \; \hat{a}_{m+1} \hat{b}_{m} C
  = 0  \; .
\label{3.21b}
\end{eqnarray}

\noindent Again, it is convenient to define new variables  $f(r),
\; g(r)$:
\begin{eqnarray}
Z_{1} = {f+g\over 2}\; , \;\; Z_{3}= {f-g \over 2} \; , \qquad
Z_{1}+ Z_{3} = f\; ,  \;\;  Z_{1}- Z_{3} = g \; ; \label{3.22}
\end{eqnarray}

\noindent then Eqs. (\ref{3.21a}) -- (\ref{3.21b})  give
\begin{eqnarray}
(- \hat{b}_{m-1} \hat{a}_{m}    + \epsilon ^{2} -k^{2}    - M^{2}
) \; {f+g\over 2}
        \nonumber\\
    + \hat{b}_{m-1} \hat{a}_{m} \; {f-g \over 2}
    +i  \hat{b}_{m-1}\hat{a}_{m} \; G  - M \sigma\; \hat{b}_{m-1}\hat{a}_{m}\; C
  =  0  \; ,
\label{3.23a}
\\
( - \hat{a}_{m+1}\hat{b}_{m}
 + \epsilon^{2} -k^{2}   - M^{2}  ) \; {f-g \over 2}
 \nonumber\\
 +   \hat{a}_{m+1} \hat{b}_{m} {f+g\over 2}  +    i \hat{a}_{m+1} \hat{b}_{m} \;
 G
   - M \sigma  \; \hat{a}_{m+1} \hat{b}_{m} C
  = 0  \; ,
\label{3.23b}
\end{eqnarray}

\noindent After simple manipulation, from two last equations it
follows that
\begin{eqnarray}
[\; - \hat{b}_{m-1}  \hat{a}_{m}    -  \hat{a}_{m+1}  \hat{b}_{m}
+  \epsilon^{2}   - k^{2}      -   M ^{2}    \; ]\;  g
           \nonumber\\
      +  ( - \hat{b}_{m-1} \hat{a}_{m}  +   \; \hat{a}_{m+1} \hat{b}_{m}  )
      \;  (  -i G   + M \sigma \;   C ) = 0 \; ,
\nonumber \\
( - \hat{b}_{m-1}  \hat{a}_{m}       + \hat{a}_{m+1}  \hat{b}_{m}
) \;  g +
 ( \epsilon^{2}   - k^{2}      -   M ^{2}    ) f
   \nonumber\\
   + (- \hat{b}_{m-1}\hat{a}_{m} -      \hat{a}_{m+1} \hat{b}_{m}  ) \;  ( - i  G  +
     M \sigma\; C  )  = 0 \; .
\label{3.24}
\end{eqnarray}

\noindent  With the use of identities
\begin{eqnarray}
-\hat{b}_{m-1} \; \hat{a}_{m}   -  \hat{a}_{m+1} \hat{b}_{m} =
\Delta  \; , \qquad - \hat{b}_{m-1} \; \hat{a}_{m}   +
\hat{a}_{m+1} \hat{b}_{m} = 2 B  \;  \nonumber
\end{eqnarray}

\noindent  Eqs. (\ref{3.24}) can be written  as
\begin{eqnarray}
[\; \Delta   +  \epsilon^{2}   - k^{2}      -   M ^{2}    \; ]\; g
      +
        2B \;  (  -i G   + M \sigma \;   C ) = 0 \; ,
\nonumber \label{3.25a}
\\
2B  \;  g +
 ( \epsilon^{2}   - k^{2}      -   M ^{2}    ) f
   +
    \Delta \;  ( - i  G  +
     M \sigma\; C  )  = 0 \; .
\label{3.25b}
\end{eqnarray}

\noindent In turn, Eqs.  (\ref{3.13}),  (\ref{3.16}) will read (in
the new variables)
\begin{eqnarray}
(  \Delta  -k^{2} -M^{2}+  \epsilon ^{2}  ) \; F  = 0 \; ,
\nonumber \label{3.26a}
\\
( \Delta  - M^{2} )\;G =
  - i ( \epsilon^{2}-k^{2} ) \; f
+ i \sigma  M \; ( \epsilon  ^{2}-k^{2}  )\; C\; . \label{3.26b}
\end{eqnarray}

Let us collect results together
\begin{eqnarray}
(  \Delta  -k^{2} -M^{2}+  \epsilon ^{2}  ) \; F  = 0 \; ,
\label{3.27a}
\\
( \Delta  - M^{2} )\;G =
  - i ( \epsilon^{2}-k^{2} ) \; f
+ i \sigma  M \; ( \epsilon  ^{2}-k^{2}  )\; C\; , \label{3.27b}
\\
(\; \Delta   +  \epsilon^{2}   - k^{2}      -   M ^{2}    \; )\; g
      +
        2B \;  (  -i G   + M \sigma \;   C ) = 0 \; ,
\label{3.27c}
\\
2B  \;  g +
 ( \epsilon^{2}   - k^{2}      -   M ^{2}    ) f
   +
    \Delta \;  ( - i  G  +
     M \sigma\; C  )  = 0 \; .
\label{3.27d}
\end{eqnarray}

It is possible to eliminate the function $C(r)$ in the above
equation. To show how it can be done, let us  turn to  a couple of
equations in (\ref{3.7c}), containing the terms  $M \;\Phi_{1}, M
\;\Phi_{3}$, and  find the combination
\begin{eqnarray}
\hat{b}_{m-1} M \Phi_{1} +  \hat{a}_{m+1} \Phi_{3}
\nonumber\\
= i\epsilon \; \hat{b}_{m-1} E_{1} - \sigma\;
\hat{b}_{m-1}\hat{a}_{m}\; C +
 i\; \hat{b}_{m-1} \hat{a}_{m}\; H_{2} - k \; \hat{b}_{m-1} H_{1}
\nonumber\\
+ i\epsilon \; \hat{a}_{m+1} E_{3} - \sigma\; \hat{a}_{m+1}
\hat{b}_{m} \; C- i \hat{a}_{m+1} \hat{b}_{m} \;   H_{2} + k \;
\hat{a}_{m+1} H_{3}
\nonumber\\
= i\epsilon \; ( \hat{b}_{m-1} E_{1} +  \hat{a}_{m+1} E_{3} ) -
\sigma\; ( \hat{b}_{m-1}\hat{a}_{m}\; +  \hat{a}_{m+1} \hat{b}_{m}
) \; C
\nonumber\\
+ i\;( \hat{b}_{m-1} \hat{a}_{m} -  \hat{a}_{m+1} \hat{b}_{m} \; )
H_{2} - k \;(  \hat{b}_{m-1} H_{1} - \hat{a}_{m+1} H_{3} )
\nonumber
\end{eqnarray}

\noindent from whence, with the help of the first and third equations
in (\ref{3.7c}) in the form
\begin{eqnarray}
  (\hat{b}_{m-1}\;
 E_{1}+ \hat{a}_{m+1} \; E_{3} ) =
 - i\epsilon \;\sigma C      - ik \; E_{2}- M \;\Phi_{0}   \; ,
\nonumber\\
(\hat{b}_{m-1} \; H_{1} - \hat{a}_{m+1} \;H_{3}) = \epsilon \;
E_{2} + \;k\sigma \; C +i M \;\Phi_{2}   \; , \nonumber
\end{eqnarray}

\noindent we  obtain
\begin{eqnarray}
\hat{b}_{m-1} M \Phi_{1} +  \hat{a}_{m+1} \Phi_{3}
\nonumber\\
= i\epsilon \; (  - i\epsilon \;\sigma C      - ik \; E_{2}- M
\;\Phi_{0} ) - \sigma\; ( \hat{b}_{m-1}\hat{a}_{m}\; +
\hat{a}_{m+1} \hat{b}_{m} ) \; C
\nonumber\\
+ i\;( \hat{b}_{m-1} \hat{a}_{m} -  \hat{a}_{m+1}  \hat{b}_{m} \;
) H_{2} - k \;(  \epsilon \; E_{2} + \;k\sigma \; C +i M
\;\Phi_{2} )\; . \nonumber
\end{eqnarray}

\noindent From this, after evident calculation, we arrive at
\begin{eqnarray}
\hat{b}_{m-1} M \Phi_{1} +  \hat{a}_{m+1} \Phi_{3} =
      - iM\;  (\epsilon  \;\Phi_{0} +  k  \;\Phi_{2})
      \nonumber\\
+ \sigma\;
(  - \hat{b}_{m-1}\hat{a}_{m}  -  \hat{a}_{m+1} \hat{b}_{m}
+\epsilon^{2}-k^{2}  ) \; C -2iB \;  H_{2} \label{3.28a'}
\end{eqnarray}

 Comparing    (\ref{3.9}) and  (\ref{3.28a'}), we conclude  that there exists  a linear relation
 \begin{eqnarray}
2iB \;  H_{2}(r)  = M ^{2}  C(r) \; . \label{3.29a}
\end{eqnarray}

\noindent Due to Eq. (\ref{3.7d}), it holds
\begin{eqnarray}
i\hat{b}_{m-1} \; \Phi_{1} - i\hat{a}_{m+1} \; \Phi_{3} = M
\;H_{2} \qquad \Longrightarrow \qquad
 i   g = M \;H_{2} \; ;
\label{3.29b}
\end{eqnarray}

\noindent therefore,  the function  $C(r)$  is expressed through
$g(r)$:
\begin{eqnarray}
C (r) = -{2B \over M^{3}} \; g(r) \label{3.29c}
\end{eqnarray}

The system (\ref{3.27a}) -- (\ref{3.27d}), after excluding $C(r)$,
takes the form
\begin{eqnarray}
(  \Delta  -k^{2} -M^{2}+  \epsilon ^{2}  ) \; F  = 0 \; ,
\label{3.30}
\\
( \Delta  - M^{2} )\;G =
  - i ( \epsilon^{2}-k^{2} ) \; f
- i \sigma    ( \epsilon  ^{2}-k^{2}  ) {2B \over M^{2}} \; g \; ,
\label{3.31a}
\\
\left [  \Delta   +  \epsilon^{2}   - k^{2}      -   M ^{2}  -
\sigma \;   ({2B \over M})^{2}     \right ] \;  g
          - 2iB \; G     = 0 \; ,
\label{3.31b}
\\
2B  \;  g +
 ( \epsilon^{2}   - k^{2}      -   M ^{2}    ) f
   +
    \Delta \;  ( - i  G  -
      \sigma\; {2B \over M^{2}} \; g   )  = 0 \; .
\label{3.31c}
\end{eqnarray}

The structure of the system allows to separate evident
linearly independent solution as follows
\begin{eqnarray}
f(r)=0 \; ,\qquad g(r)=0\;, \qquad H(r)=0\;,
\nonumber\\
F(r) \neq 0 \; , \qquad  (  \Delta  -k^{2} -M^{2}+  \epsilon ^{2}
) \; F  = 0 \; ;\label{3.32}
\end{eqnarray}

\noindent corresponding functions and the energy spectrum are known
(also see below).

We are to solve the system of three  last
equations in   (\ref{3.31a}) -- (\ref{3.31c}). With the help of
(\ref{3.31b})
\begin{eqnarray}
 \Delta  \;G
=  M^{2} G   - i ( \epsilon^{2}-k^{2} ) \; f - i \sigma   \; (
\epsilon  ^{2}-k^{2}  ) {2B \over M^{2}} \; g \; ,
\nonumber\\
   \Delta \; g =  -[ \epsilon^{2}   -
k^{2}      -   M ^{2}  -  \sigma    ({2B \over M})^{2} ] \;  g
          - 2iB \; G      \; ,
\nonumber
\end{eqnarray}

\noindent Eq.  (\ref{3.31c}) takes the form of the linear relation
\begin{eqnarray}
   M ^{2}     f
 = i\left ( -  M^{2}
 +   \sigma  {4B^{2} \over M^{2}}  \right ) \; G +
 2B  \left ( 1     - \sigma - { \sigma^{2} \over M^{2} }   {4B^{2} \over M^{2}}   \right )    g\; .
\label{3.33}
\end{eqnarray}

\noindent  Now, returning to Eqs.  (\ref{3.31a})--(\ref{3.31b}),
after excluding the function
 $f$ and using the notation
\begin{eqnarray}
\gamma ={\epsilon^{2}- k^{2} \over M^{2}} \; , \qquad \beta =
\sigma {4B^{2} \over M^{2}}\;, \qquad \alpha = \gamma \; \rho ,
\qquad \rho = 1 - { 4B^{2}\sigma^{2} \over M^{4} }    \; ;
\nonumber
\end{eqnarray}

\noindent we arrive at  two equations
\begin{eqnarray}
 (  \Delta   +  \epsilon^{2}   - k^{2}      -   M ^{2}    ) \;  g=
      \beta  \; g (r)       + 2iB \; G (r)     \; ,
\nonumber
\\
( \Delta + \epsilon^{2}- k^{2} - M^{2} )\;G =
   -  2iB \alpha  \;  g(r)
     +       \beta   \gamma  \; G(r)
   \; .
\label{3.34}
\end{eqnarray}

In matrix form they read
\begin{eqnarray}
( \Delta    + \epsilon^{2}    - M^{2} -   k ^{2}   ) \left |
\begin{array}{c}
g (r)\\
G(r)
\end{array} \right | =
\left | \begin{array}{cc}
  \beta    & 2iB \\
-2iB \alpha  &   \beta   \gamma
\end{array} \right |
  \left | \begin{array}{c}
g(r) \\
G(r)
\end{array} \right | \; .
\label{3.35}
\end{eqnarray}

\noindent Let us  construct the transformation changing the matrix on
the right to a diagonal form
\begin{eqnarray}
( \Delta    + \epsilon^{2}    - M^{2} -   k ^{2}   )  \left |
\begin{array}{c}
g '\\
G '
\end{array} \right | =
\left | \begin{array}{cc}
\lambda_{1}   & 0 \\
0   & \lambda_{2}
\end{array} \right |
  \left | \begin{array}{c}
g '\\
G'
\end{array} \right |,
\nonumber\\
\left | \begin{array}{c}
g '\\
G'
\end{array} \right | = S \; \left | \begin{array}{c}
g \\
G
\end{array} \right | \; , \qquad
S = \left | \begin{array}{cc}
s_{11} & s_{12} \\
s_{21} & s_{22}
\end{array} \right | ;
\label{3.36}
\end{eqnarray}

\noindent the  problem is reduced to a couple of linear systems
\begin{eqnarray}
\left \{ \begin{array}{l}
(\beta  - \lambda_{1} ) \; s_{11} \;   -  2iB\alpha \; s_{12}  =  0 \;, \\[2mm]
 2iB \; s_{11}  + (\beta \; \gamma - \lambda_{1})\; _{12} \;  = 0 \;,
\end{array} \right. \qquad
\left \{ \begin{array}{l}
 ( \beta -\lambda_{2} ) \;  s_{21}   -  2iB\alpha \; s_{22}  = 0  \; ,\\[2mm]
    2iB\; s_{21}   +  (\beta \gamma   - \lambda_{2} )\; s_{22}  = 0  \; .
  \end{array} \right.
\nonumber \label{3.37b}
\end{eqnarray}

\noindent
The eigenvalues  $\lambda_{1}, \lambda_{2}$  are
\begin{eqnarray}
\lambda_{1} = {\beta (1+\gamma) +  \sqrt{ \beta^{2}
(1-\gamma)^{2} + 16B^{2} \rho \gamma   }\over 2}\; , \nonumber
\\
 \lambda_{2} = {\beta(1+\gamma) -
 \sqrt{\beta^{2}(1-\gamma)^{2} + 16 B^{2}\rho \gamma    }\over 2}
 \; . \label{3.39}
\end{eqnarray}

\noindent let it be
\begin{eqnarray}
s_{12}=1,  \; s_{22}=1 \;, \qquad s_{11} =  {\lambda_{1}' - \beta
\gamma \over 2i B} \; , \;\; \qquad
  s_{21} = {\lambda_{2} - \beta \gamma \over 2i B}\; ,
\nonumber\\
g' = {\lambda_{1} - \beta \gamma \over 2i B}  g + G \;, \qquad G'
=
 {\lambda_{2} - \beta \gamma \over 2i B}  g + G \; ;
\label{3.40b}
\end{eqnarray}

In the new (primed) basis, Eqs.  (\ref{3.36})    take the form of
two separated differential equations
\begin{eqnarray}
  \left (  \; \Delta   +  \epsilon^{2}   - k^{2}      - M
^{2}  -  \lambda_{1}   \;  \right  )  \; g' =0 \; ,
\nonumber
\\
   \left ( \;  \Delta   +  \epsilon^{2}   - k^{2}      -   M ^{2}  -  \lambda_{2}'  \;  \right  )  \; G' =0 \; ;
\label{3.42a}
\end{eqnarray}

Recalling the meaning of $\Delta$, let us specify the second order
equation
\begin{eqnarray}
   \left (
   {d^{2} \over dr^{2}}  + {1 \over r}{d \over d r} -
{(m+Br^{2})^{2} \over r^{2}}     + \lambda^{2}  \right  ) \varphi
(r) =0 \; . \label{3.43}
\end{eqnarray}

\noindent This equation was examined above.
We   obtain two
possibilities for the energy spectrum:
\begin{eqnarray}
g' \neq 0\; ,  \qquad \epsilon^{2}   -
k^{2}  = \lambda^{2}  +    M ^{2}  + \lambda_{1} \; , \nonumber
\\
 G' \neq 0 \; , \qquad  \epsilon^{2}   -
k^{2}  = \lambda^{2}  +    M ^{2}  + \lambda_{2}'  \; .
\label{3.46}
\end{eqnarray}

\noindent Both Eqs. (\ref{3.46}) can be written  as
\begin{eqnarray}
M^{2} \gamma   = \lambda^{2}  +    M ^{2}+ {\beta(1+\gamma) \pm
\sqrt{\beta^{2} (1-\gamma)^{2} + 16B^{2}\rho  \gamma   }\over 2}
\; . \label{3.47}
\end{eqnarray}

It is convenient to introduce  new  variable    $x = \gamma -1
$, and also with the help of
\begin{eqnarray}
 \beta = \sigma {4B^{2} \over M^{2}}\;, \qquad \rho = 1 - {
4B^{2}\sigma^{2} \over M^{4} }     = 1 - {\beta^{2} \over 4B^{2}}
\; , \qquad 16 \rho B^{2} = 16 B^{2} - 4 \beta^{2} \nonumber
\end{eqnarray}

\noindent to eliminate the parameter  $\rho$:
\begin{eqnarray}
(2M^{2} -\beta) \; x   - 2 (\lambda^{2} + \beta )  = \pm
\sqrt{\beta^{2} x^{2} + (16 B^{2} - 4 \beta^{2} ) (x+1) } \; .
\label{3.49}
\end{eqnarray}

\noindent Thus, we get the second  order  equation
\begin{eqnarray}
M^{2}(M^{2} - \beta)    x^{2} -  \left [ (\lambda^{2} +
\beta)(2M^{2}-\beta) + (4B^{2} -  \beta^{2} ) \right ] x
\nonumber \\
+  (\lambda^{2} + \beta)^{2}  - (4B^{2} -  \beta^{2} ) = 0 \; ;
\label{3.50}
\end{eqnarray}

\noindent   its
solutions read
\begin{eqnarray}
\epsilon^{2} - M^{2} - k^{2} =
 {1 \over 2 (M^{2} - \beta)   } \left \{ \;  [ \;
(\lambda^{2} + \beta)(2M^{2}-\beta) + (4B^{2} -  \beta^{2} ) \;  ]
 \right. \nonumber
\\
\left.  \pm  \left [   [ \; (\lambda^{2} + \beta)(2M^{2}-\beta) +
(4B^{2} -  \beta^{2} ) \;  ]^{2}   \right. \right.
\nonumber
\\
\left. \left.  -4 M^{2}(M^{2} - \beta)
(\lambda^{2} + \beta)^{2}  - (4B^{2} -  \beta^{2} ) \right ]  \right \} \;
.
\label{3.51}
\end{eqnarray}

Note, that the case (\ref{3.32}) gives  the following spectrum
$
\epsilon^{2} = M^{2} + k^{2} + \lambda^{2} $,
 so for  these solutions the polarizability does not manifest itself in a magnetic field.

Thus,  on the base of  general covariant formalism
 in   the vector particle with polarizability,  the exact solutions for such a particle  are constructed
 in the presence of an external homogeneous magnetic
 field.
 There are separated three types of linearly independent solutions, and corresponding energy spectra are found.

The authors are grateful to the participants of the seminar of
the Laboratory of Theoretical Physics, Institute of Physics, National
Academy of Sciences of Belarus, for stimulating discussion.

\end{document}